# Duality, Mechanical Wave Theory, New Quantum Operators and Nonlinear Equations in Quantum Theory

Yi-Fang Chang

Department of Physics, Yunnan University, Kunming 650091, China

(E-mail: yifangchang1030@hotmail.com)

**Abstract**: Various dualities are summarized. Based on the universal wave-particle duality, along an opposite direction of the developed quantum mechanics, we use a method where the wave quantities frequency v and wave length $\lambda$ are replaced on various mechanical equations, and may be derive some new results. It is called the mechanical wave theory. From this we derive new operators which represent more physical quantities. Further, we propose some nonlinear equations and their solutions, which may be probably applied to quantum theory.

## 1. Introduce

The wave-particle duality is very important basis of quantum mechanics. Though Heisenberg insisted that one could interpret the quantum-mechanical equation of motion in terms of either a wave ontology or a particle ontology, and Bohr emphasized wider wave–particle complementarity.

Quantum theory predicts that proposed measurements on the amplified field will indicate that the single input photon exhibits extremes of both wavelike and particlelike characteristics. Swanson and Carlsten analyzed theoretically particlelike and wavelike characteristics of single input photons by utilizing amplifiers in a modified interferometer, and the evidence for wavelike behavior is fringe visibility at the output of the interferometer [1]. Anticorrelation between the outputs of the two legs of the interferometer provides the evidence for particlelike behavior. Although quantum noise degrades both the particle-sensitive and the wave-sensitive measurement, no trade-off between the wavelike and the particlelike characteristics takes place. Complementarity is maintained by the quantum noise added during amplification [1]. Englert derived an inequality, which can be regarded as quantifying the notion of wave-particle duality, by the fringe visibility in a two-way interferometer sets an absolute upper bound on the amount of which-way information that is potentially stored in a which-way detector [2]. Naschie analyzed the wave-particle duality from the geometry and the topology of micro spacetime, and shown that the index theorem of Toepliz operators is consistent with the irreducible spacial uncertainty of Cantorian spacetime and that this is responsible for various paradoxial notions in quantum physics, such as the wave-particle duality [3]. Schwindt, et al., quantified the wave-particle duality by the inequality, and investigated the relation for various situations, including pure, mixed, and partially mixed input states [4]. Abranyos, et al., proposed for the experimental verification of an inequality quantified the concept of wave-particle duality relating (or setting) an upper limit on distinguishability and visibility in a two-way interferometer [5]. Van Meter, et al., discussed the radiative corrections in symmetrized classical electrodynamics, and outlined the connection with electromagnetic duality, and given an in-depth discussion of nonlocal four-momentum conservation for the wave-particle system [6]. Ryff and Ribeiro proposed an experiment that permits observation of the de Broglie two-photon wave packet behavior for a pair of photons by a Mach-Zehnder interferometer, and the same technique can give which-path information via an interaction-free experiment and can be used in other experiments on the foundations of quantum mechanics related to wave-particle duality and to nonlocality [7]. Durr proposed quantitative measures of wave properties and particle properties for multibeam interferometers, and shown that these quantities are connected by a few fundamental inequalities which express wave-particle duality [8]. Wiseman discussed the quantum-trajectory theory, which is applied to the cavity-QED experiment demonstrating wave particle duality [9]. Gatti, et al., formulated a theory for entangled imaging, and shown that the results for imaging and for the wave-particle duality features [10]. Hara and Ohba calculated a tunneling time distribution by means of Nelson quantum mechanics and investigated the wave-particle duality [11]. Yang discussed the wave-particle duality in complex space [12]. Utter and Feagin analyzed the trapped-ion realization of Einstein historic recoiling-slit gedanken experiment, and quantified the photon-path which-port information cached in the recoiling ion and the underlying wave-particle duality [13]. Miniatura, et al., discussed that the wave-particle duality is applied to electrons or light propagating in disordered media [14]. Schmidt, et al., reported the evidence of wave-particle duality for single 1.3 MeV fast hydrogen atoms formed close to either target nucleus in electron-transfer collisions [15]. Liu, et al., researched the wave-particle duality relation and joint measurement in a Mach-Zehnder interferometer [16]. The wave-particle duality of massive objects is a cornerstone of quantum physics and a key property of many modern tools such as electron microscopy, neutron diffraction or atom interferometry. Juffmann, et al., reported the experimental demonstration of quantum interference lithography with complex molecules [17]. Haan, et al., provided the experimental determination of the Goos-Hanchen shift for a particle experiencing a potential well as required by wave-particle duality [18].

Based on the different characteristic at low and high energies in particle, we suggested the symmetry-statistics duality, which is possible development of the wave-particle duality. The present theories and models, etc., of particle seem to be divided into symmetrical and statistical [19].

Recently, the gauge-gravity duality is investigated. Klebanov, et al., studied effects associated with the chiral anomaly for a cascading gauge theory using gauge-gravity duality [20]. Herzog discussed gauge-gravity duality, and string tensions and confining (2+1)-dimensional gauge theories [21]. Buchel, et al., studied time-dependent solutions of supergravity and string theory, and obtained analytical continuation of static solutions and exactly solvable worldsheet models from de Sitter deformation of gauge-gravity dualities [22]. Son and Stephanov considered the model of an open moose by phenomenological models of hidden local symmetries and the ideas of dimensional deconstruction and gauge-gravity duality [23]. Kovtun and Starinets used the gauge-gravity duality conjecture to compute spectral functions of the stress-energy tensor in finite-temperature N=4 supersymmetric Yang-Mills theory [24]. Janik and Peschanski derived the equation for the quasinormal modes corresponding to the scalar excitation of a black hole moving away in the fifth dimension, and discussed gauge-gravity duality and thermalization of a boost-invariant expanding perfect fluid in N=4 SUSY Yang-Mills theory at large proper-time [25]. Hanada, et al., proposed circumventing all these problems inherent in the lattice approach by adopting a nonlattice approach for one-dimensional supersymmetric gauge theories, which can be used to investigate the gauge-gravity duality from first principles, and to simulate M theory based on the matrix theory conjecture [26]. Anagnostopoulos, et al., presented the first Monte Carlo results for supersymmetric matrix quantum mechanics with 16 supercharges at finite temperature, and these results provide highly nontrivial evidence for the gauge-gravity duality [27]. Erdmenger, et al., considered gauge-gravity duality with flavor for the finite-temperature field theory dual of the AdS-Schwarzschild black hole background with embedded D7-brane probes, and investigated the holographic vector mesons from spectral functions at finite baryon or isospin density [28]. Skenderis and van Rees investigated the real-time gauge-gravity duality [29]. Argurio, et al., considered the interplay of various fractional branes and the corresponding gauge-gravity duality [30]. Ishiki, et al., studied the deconfinement phase transition in N=4 super Yang-Mills theory from supersymmetric matrix quantum mechanics, which corresponds to the Hawking-Page transition based on the gauge-gravity duality [31]. Hanada, et al., performed a direct test of the gauge-gravity duality associated with the system in type IIA superstring theory at finite temperature, and derived corrections to the black hole thermodynamics from supersymmetric matrix quantum mechanics [32]. Using gauge-gravity duality, Chesler and Yaffe studied the creation and evolution of anisotropic, homogeneous strongly coupled N=4 supersymmetric Yang-Mills plasma, which corresponds to horizon formation and far-from-equilibrium isotropization in this plasma [33]. Zhou proposed a Dirac-Born-Infeld vertex fundamental strings configuration for a probe baryon in the finite-temperature thermal gauge field, and discussed D4 brane probes in gauge-gravity duality [34]. Nogueira derived an exact formula for the central charge of the U(1) current in terms of the gauge coupling at quantum criticality and compare it with the corresponding result obtained using gauge-gravity duality. Here the amplitude of the current correlation function has the same form as predicted by the gauge-gravity duality, and compared finite temperature results for the charge susceptibility with the result predicted by the gauge-gravity duality [35]. Alanen, et al., computed the spatial string tension of finite temperature QCD matter in gauge-gravity duality [36], and studied a gauge-gravity duality model for the thermodynamics of a gauge theory with one running coupling [37]. Then they used gauge-gravity duality to study the thermodynamics of a field theory with asymptotic freedom in the ultraviolet and a fixed point in the infrared [38]. McGuirk, et al., considered a relative of semidirect gauge mediation, and construction of holographic gauge mediation, which are amenable to the techniques of gauge-gravity duality [39]. Brodsky, et al., discussed a nonperturbative effective coupling from the light-front holographic mapping of classical gravity, which gives support to the application of the gauge-gravity duality to the confining dynamics of strongly coupled QCD [40]. Using gauge-gravity duality, Athanasiou, et al., computed the energy density and angular distribution of the power radiated in strongly coupled N=4 supersymmetric Yang-Mills theory, and discussed synchrotron radiation in the strongly coupled conformal field theories [41]; Chesler and Yaffe studied the creation and evolution of boost-invariant anisotropic, strongly-coupled N=4 supersymmetric Yang-Mills plasma [42].

Moreover, other dualities are also discussed. Recently, Karch investigated the action of the SL(2,Z) electric-magnetic duality group and topological insulators, and used electric-magnetic duality to find a gravitational dual for a strongly coupled version of this theory using the gauge-gravity correspondence [43]. He and Miao studied the special case of SU(2) pure gauge theory, and obtained the magnetic and dyonic expansions of the Nekrasov theory, and discussed the relation between the electric-magnetic duality of gauge theory and the action-action duality of the integrable system [44]. Cobanera, et al., researched unified approach to classical and quantum dualities, and duality relations appear only in a sector of certain theories (emergent dualities), and obtained new self-dualities for four-dimensional Abelian gauge field theories [45]. Biswas, et al., discussed the nature of the Hagedorn transition and a thermal duality [46].

Malace, et al., investigated the neutron structure function and the confirmation of quark-hadron duality [47]. González-Alonso, et al., assessed the uncertainties associated with the violations of quark-hadron duality [48], and analyzed the pinched weights which are thought to reduce the violation of quark-hadron duality in finite-energy sum rules [49]. Goldstein and Berkovits proved an exact duality between different geometries of a resonant level in a Luttinger liquid [50]. Girardeau and Astrakharchik found analytically the ground-state wave function by a Bose-Bose duality mapping [51]. Ranninger and Domanski discussed the boson-fermion duality and the intrinsic structural metastability in cuprate superconductors [52].

Using the gauge-string duality, Noronha shown that in four-dimensional gauge theories dual to five-dimensional Einstein gravity coupled to a single scalar field, which connects with the Polyakov loop to the thermodynamics of SU(N) gauge theories [53]; Andreev estimated the quadratic correction in the pressure of cold quark matter [54]; Gao and Mou investigated deep inelastic scattering off the polarized neutron [55]. Argurio and Dehouck studied that gravitational duality acts on rotating solutions [56]. Shifman and Yung observed a crossover transition from weak to strong coupling in N=2 supersymmetric QCD, and the non-Abelian bulk duality is in one-to-one correspondence with a duality taking place in the supersymmetric model [57].

Kuzenko discussed the Fayet-Iliopoulos term for general models for self-dual nonlinear supersymmetric electrodynamics, and presented a two-parameter duality-covariant deformation of the N=1 supersymmetric Born-Infeld action as a model for partial breaking of N=2 supersymmetry [58]. Bourliot, et al., discussed four-dimensional spatially homogeneous gravitational instantons, self-duality and geometric flow [59]. Xu and Horava proposed a model with a stable algebraic Bose liquid phase at low energy guaranteed by the gauge symmetry of gravitons and self-duality of the low-energy field theory [60]. Tamaryan, et al., used this duality between unit vectors and highly entangled W states to find the geometric measure of entanglement of such states [61].

By introducing a generalized distinguishability as a measure of the unsharpness of an unsharp measurement, Yu, et al., derived a complementarity inequality, which generalizes Englert duality inequality, from the condition of joint measurement of two orthogonal unsharp observables [62]. Deconinck and Terhal used the quantum state discrimination duality theorem, and discussed the qubit state discrimination [63]. Megías et al., analyzed the duality between QCD perturbative series and power corrections [64]. Boer and Shigemori used the U-duality symmetry to predict exotic branes and nongeometric backgrounds [65]. Seidel discussed the S-duality constraints on 1D patterns associated with fractional quantum Hall states [66]. Via a duality between entanglement and classical correlations, Adesso and Datta derived a closed formula for the Gaussian entanglement of formation of a family of three-mode Gaussian states [67]. Akhanjee and Rudnick discussed spherical spin-glass-Coulomb-gas duality, and presented a solution beyond mean-field theory [68]. Bern, et al., observed that classical tree-level gauge-theory amplitudes can be rearranged to display a duality between color and kinematics, and this duality persists to all quantum loop orders, and the three-loop four-point amplitude of N=4 super-Yang-Mills theory can be arranged into a form satisfying the duality [69]. Using the superconformal indices techniques, Spiridonov and Vartanov constructed Seiberg type dualities for N=1 supersymmetric field theories outside the conformal windows [70]. Kim, et al., discussed the duality between the space-dependent and time-dependent electric fields of the same form at the leading order of the effective actions of QED [71].

**2. The mechanical wave theory**

It is well-known that quantum mechanics is based on the microscopic wave-particle duality. From Planck and de Broglie relations:

$$E = mc^2 = h\nu, \quad p = mv = (h\lambda^{-1})k_0 = h\vec{k}, \quad (1)$$

and a developed similarity between geometrical optics and classical mechanics found by Hamilton in 19 century, Schrodinger used the frequency $v = E/p = E/\sqrt{2m(E-V)}$, and the variables energy E and momentum p in the classical relation are replaced by the operators $\hat{E}$ and $\hat{p}$ on the general wave equation:

$$\nabla^2 \varphi - \frac{1}{u^2}\frac{\partial^2 \varphi}{\partial t^2} = 0. \quad (2)$$

Then he derived a known Schrodinger equation in quantum mechanics [72,73]:

$$\nabla^2 \psi + \frac{2m}{\hbar^2}(E-V)\psi = 0. \quad (3)$$

This is a result that the physical quantities E and p are replaced on the wave equation.

For the energy-momentum relation

$$E^2 = p^2c^2 + m^2c^4, \quad (4)$$

the variables E and p in the relativity are replaced by the operators $\hat{E}$ and $\hat{p}$, and then the Klein-Gordon equation is derived [74].

Along an opposite direction of the developed quantum mechanics, we used a method where the wave quantities frequency v and wave length $\lambda$ are replaced on various mechanical equations in classical mechanics and relativity, and then obtain some new results. It is also based on the microscopic wave-particle duality, and should be called the mechanical wave theory [75,76]. This includes: the Newtonian equation of motion:

$$F=dP/dt \rightarrow \text{(developed to)} \; F = \hbar d(\lambda^{-1}k_0)/dt = \hbar d\vec{k}/dt. \quad (5)$$

From this the change rate of the wave vector or wave length and its direction with time are direct ratio with force. This may extend to the four-dimensional generalized form:

$$F_s = \hbar dk_s / dt. \quad (6)$$

The corresponding Newtonian gravitational law is:

$$F = -G\frac{m_1 m_2}{r^2}\vec{r}_0 \rightarrow F = -\frac{Gh^2}{c^4}\frac{\nu_1 \nu_2}{r^2}\vec{r}_0 \quad (7)$$

According to this law there should have gravitation between both waves and their frequencies. The potential of wave may be similarly derived

$$\varphi = -Gh\nu/c^2 r. \quad (8)$$

In the mechanical wave theory, if the quantization of the wave-particle duality is not considered, this will recur to a classical wave theory for h→0. It shows that in this case the forces and potentials of interactions cannot affect any wave. For a set of the wave the momentum is:

$$p = \sum m_i v_i \to p = h \sum (\lambda_i^{-1} k_{0i}) = h \sum k_i . \tag{9}$$

The Hamilton-Jacobi equations are:

$$p_\alpha = -\frac{\partial S}{\partial x_\alpha} \to k_\alpha = -\frac{1}{\hbar}\frac{\partial S}{\partial x_\alpha} . \tag{10}$$

Correspondingly, the four-dimensional generalized form of the Hamilton-Jacobi equation is:

$$k_s = -\frac{1}{\hbar}\frac{\partial S}{\partial q_s} , \tag{11}$$

The canonical equations of Hamilton are:

$$\frac{\partial H}{\partial p} = \dot{q} \to \frac{dv}{dk} = \frac{dx}{dt} , \tag{12}$$

which shows that the group velocity equals the particle velocity; and

$$\frac{\partial H}{\partial q} = -\dot{p} \to \frac{\partial v}{\partial x} = -\frac{dk}{dt} , \tag{13}$$

which shows that the change rate of wave frequency with space is direct ratio with force. Conversely, when the force is zero, k, $\lambda$ or v are the movement integral. An object has oscillation, which corresponds to wave, both equations are parallelism each other.

In relativistic mechanics, the formula of energy and momentum of an object with rest mass, and the corresponding frequency and wave-vector is:

$$E = \frac{m_0 c^2}{\sqrt{1-\beta^2}} (\beta = \frac{u}{c}) \to v = \frac{v_0}{\sqrt{1-\beta^2}} . \tag{14}$$

It is namely the Doppler transverse effect.

$$p = \frac{m_0 v}{\sqrt{1-\beta^2}} \to k = \frac{k_0}{\sqrt{1-\beta^2}} , \tag{15}$$

The frequency and wave-vector increase along with increase of movement of the wave-sources. While the general wave equation is:

$$\Delta \varphi - \frac{1}{u^2}\frac{\partial^2 \varphi}{\partial t^2} = 0 , \tag{16}$$

which shows the constancy of the wave-velocity [77,19].

$$m^2 c^4 = m_0^2 c^4 + c^2 p^2 \to v^2 = v_0^2 + c^2 k^2 = v_0^2 (1 + c^2 k^2 / v_0^2) . \tag{17}$$

The energy-momentum tensor of the simplest macroscopic object is [77]:

$$T_{ik} = mc^2 u_i u_k , \tag{18}$$

which extends to wave, so we obtain the frequency-wave vector tensor:

$$T_{ik} = h\nu u_i u_k.  \tag{19}$$

Here $-T_{44} = h\nu$ is energy, $-(i/c)T_{\alpha 4} = (h\nu/c)(v/c) = hk$ is momentum, $-icT_{\alpha 4} = hw_\alpha$ is energy flux and $T_{\alpha\beta} = hk_\alpha v_\beta$ is momentum flux.

$$\frac{dp_i}{ds} = -\frac{\partial T_{ik}}{\partial x_k} \to \frac{dk_i}{ds} = -\frac{\partial(\nu u_i u_k)}{\partial x_k}. \tag{20}$$

In the gravitational field

$$E = m_0 c^2 - m_0\varphi \to h\nu = h\nu_0 - h\nu_0 \varphi/c^2, \tag{21}$$

$$\therefore \nu = \nu_0(1-\varphi/c^2). \tag{22}$$

It shows that the frequency changes in gravitational field, i.e., the gravitational red shift of the general wave. According to general relativity, any wave should also produce a gravitational field with the mass $h\nu/c^2$, and derive the curved space-time. This is namely the wave field with curved space-time. In the weak field

$$g_{00} = 1 + \gamma_{00}, \text{ here } \gamma_{00} = 2\varphi/c^2 = -2Gm/c^2 r = -2Gh\nu/c^4 r, \tag{23}$$

$$R_0^0 \approx \frac{1}{c^2}\Delta\varphi. \tag{24}$$

Above some equations are consistent with Heisenberg equation:

$$\frac{d\hat{F}}{dt} = \frac{\partial \hat{F}}{\partial t} + [\hat{F}, \hat{H}], \tag{25}$$

and the Ehrenfest law:

$$\hat{F} = \frac{d\hat{p}}{dt}, \hat{v} = \frac{1}{m}\hat{p}. \tag{26}$$

In a word, all wave which holds for the wave-particle duality may apply various equations, theories, methods and formulas on classical particle, relativity and quantum mechanics, for example, it should includes light and sound wave, etc.

### 3. New possible operators

When various wave quantities are replaced on quantum mechanics, we may obtain various operators of the corresponding quantities. This is also an extension of operator representation. For instance,

$$p_\mu = -i\hbar\frac{\partial}{\partial x_\mu}, \tag{27}$$

which includes

$$p_\alpha = -i\hbar \frac{\partial}{\partial x_\alpha}, E = i\hbar \frac{\partial}{\partial t}. \tag{28}$$

The four-dimensional wave vector operators and four-dimensional velocity operators are:

$$k_\mu = \frac{p_\mu}{\hbar} = -i\frac{\partial}{\partial x_\mu}, \quad v_\mu = \frac{p_\mu}{m} = -i\frac{\hbar}{m}\frac{\partial}{\partial x_\mu}. \tag{29}$$

The frequency operator and wave vector operator extends to the four-dimensional generalized form:

$$k_s = \frac{p_s}{\hbar} = -i\frac{\partial}{\partial q_s}. \tag{30}$$

The four-dimensional generalized operator is:

$$v_s = \frac{p_s}{m} = -i\frac{\hbar}{m}\frac{\partial}{\partial q_s}. \tag{31}$$

The force operator is:

$$F = \frac{dp}{dt} = -i\hbar \frac{d}{dt}\nabla. \tag{32}$$

$$\because p_\mu = \frac{\partial S}{\partial x_\mu}, \tag{33}$$

so the operator of Hamilton-Jacobi function is:

$$S = -i\hbar. \tag{34}$$

The operator of phase (eikonal) function [77] is:

$$\theta = -i. \tag{35}$$

Therefore, various physical quantities (force, velocity and so on) and wave should all be quantized. Wave as field relates is again the quantization of field.

In relativistic mechanics the Hamilton-Jacobi equation is:

$$(\frac{\partial S}{\partial x_i})^2 = -m^2c^2 \to -\hbar^2 \frac{\partial^2}{\partial x_i^2} = -m^2c^2. \tag{36}$$

This corresponds to the relativistic Klein-Gordon equation:

$$(\Box - \frac{m^2c^2}{\hbar^2})\Psi = 0. \tag{37}$$

The classical Hamilton-Jacobi equation is:

$$\frac{1}{2m}(\frac{\partial S'}{\partial x_\alpha})^2 = -\frac{\partial S'}{\partial t} \to -\frac{\hbar^2}{2m}\nabla^2\Psi = i\hbar \frac{\partial \Psi}{\partial t}, \tag{38}$$

This corresponds to the nonlinear Schrodinger equation.

The uncertainty relation [78] is:

$$\Delta x_i \Delta p_i = \hbar. \tag{39}$$

It corresponds to the band-width law of wave [77], i.e., the uncertainty relation of wave, which includes:

$$\Delta t \Delta \nu \approx 1, \quad \Delta x \Delta(\lambda^{-1}) \approx 1. \tag{40}$$

It extends to the four-dimensional generalized form:

$$\Delta x_i \Delta k_i \approx 1. \tag{41}$$

This shows that the Heisenberg uncertainty relation for particles with wave and duality is only an extension of above property on wave. Simultaneously, when mass is constant, from operator of velocity we may obtain the uncertainty relations of velocity:

$$\Delta v_\mu \Delta x_\mu = \hbar/m, \quad \Delta v_\mu \Delta x_\mu = \hbar/m. \tag{42}$$

They are the same with our results [79]. From the uncertainty relations of velocity the speed of light as the velocity will be uncertain, and the speed of light should have the statistical fluctuations inside a small space-time and at high energy. These results are contradiction with the constancy of the vacuum speed of light, which is usually considered as a constant for quantum theory. We believe that relativity and quantum theory will be able to be unified completely only after both are developed [79]. The principle of least action corresponds to the Fermat principle of geometrical optics, and to the path integral of quantum mechanics.

## 4. New nonlinear quantum equations and their some solutions

If the wave-particle duality holds always, the wave property of particle will be necessarily dispersion according to quantum mechanics. Such only it combines the nonlinearity, a particle may be stable. In this case the particle is namely a soliton, which corresponds to the nonlinear wave, and nonlinear operator and nonlinear equation [19,80]. Combining the physical fact (stability) and the mathematical logic, quantum mechanics and quantum theory must be nonlinear under these conditions. Except various known nonlinear Schrodinger equation, nonlinear Dirac equations, sine-Gordon equation and nonlinear Klein-Gordon equation, since Schrodinger equation with a potential V=0 is:

$$\frac{\partial \psi}{\partial t} = \frac{i\hbar}{2m} \nabla^2 \psi, \tag{43}$$

which is analogue with the diffusion equation:

$$\frac{\partial u}{\partial t} = \mu \nabla^2 u. \tag{44}$$

Therefore, the diffusion equation is developed to the nonlinear Burgers equation, the KdV equation, etc., Schrodinger equation should be able to develop to following nonlinear equations:

$$\frac{\partial \psi}{\partial t} + \alpha \psi \nabla \psi = \frac{i\hbar}{2m} \nabla^2 \psi; \tag{45}$$

and a similar KdV equation:

$$\frac{\partial \psi}{\partial t} + \sigma \psi \frac{\partial \psi}{\partial x} = \frac{\partial^3 \psi}{\partial x^3}. \tag{46}$$

Its solution may represent N solitons proceeding with the velocity corresponding to the N discrete eigenvalues [81-83]. A solution of KdV equation is [81]:

$$\psi = (\frac{72}{\sigma})\frac{3+4ch(2x-8t)+ch(4x-64t)}{[3ch(x-28t)+ch(3x-36t)]^2}. \qquad (47)$$

At the same time, from Boussinesq equation and Klein-Gordon equation we may develop similarly to the nonlinear equation:

$$\phi_{tt} - \phi_{xx} = \sigma(\phi^2)_{xx} + \phi_{xxxx}. \qquad (48)$$

For this a solution is [83]:

$$\phi(x,t) = \frac{1}{4}k^2 \sec h^2[\frac{1}{2}(kx - \beta t + \delta)]. \qquad (49)$$

Here $\beta^2 = k^2 + k^4$.

Conversely, the wave quantities v and $\lambda$ may be replaced on the nonlinear mechanical equation. For example, in hydrodynamics the known Navier-Stokes equation is:

$$\rho \frac{\partial \vec{u}}{\partial t} + \rho \vec{u} div \vec{u} = \vec{F} - grad p + \mu \Delta \vec{u}. \qquad (50)$$

This is already applied to a nonlinear dynamical formation model on binary stars [84,85]. And the equation is

$$m\frac{\partial v}{\partial t} + \nabla(\frac{1}{2}mv^2) - (mv \times rot v) = F - \frac{m}{\rho} grad p. \qquad (51)$$

Suppose that the force in liquid possesses potential, so $F = -\nabla U$. When this is extended to wave, since the wave velocity is changeable, and wave has kinetic energy, and has also potential energy, if $(mv^2/2) + U = E = h\nu$, the equation (51) will become:

$$h\frac{\partial k}{\partial t} + h\nabla v = hk \times rot v - V grad p. \qquad (52)$$

Here $V = h\nu/c^2 \rho$ is a volume. The known light pressure is $p = (1+\gamma)Nh\nu/c$, here $\gamma$ is a reflectance, and then a complete similar wave pressure should be:

$$p = (1+\gamma)P. \qquad (53)$$

Therefore, Eq.(52) becomes:

$$\frac{\partial k}{\partial t} + \nabla v = k \times rot v - \frac{h\nu}{\rho c^2}(1+\gamma) grad |k|. \qquad (54)$$

Perhaps, this may be called the hydrodynamics of wave. For the irrotational wave, rot**v**=**0**.

Various nonlinear waves and their equations may be applied to quantum theory and particle physics. We proposed some nonlinear equations with double solutions of soliton and chaos, which corresponds possibly to the wave-particle duality in quantum theory, and possess some new meanings in mathematics, physics and particle theory [86]. The figure of Eq.(47) [81] is very analogue with some elastic scatterings of particles.

## 5. Summary

Based on the analysis of the logical structure of quantum mechanics, we think that the wave-particle duality is the only basic principle of quantum mechanics. Statistics is the corresponding mathematical characteristic. The other principles, for example, the superposition principle, the operator representations of dynamical variables, the uncertainty principle, the eigenvalue equations, the identical principle and so on [73,78], are all physical or mathematical results [19].

In a word, no matter what the physical quantities E and p of particles are replaced on the nonlinear wave equations, or the physical quantities v and $\lambda$ of wave are replaced on the nonlinear mechanics equations, we may all derive the nonlinear quantum mechanics [19,80]. Both possess symmetry.

We pointed out that the present applied superposition principle is linear, it must be developed into a generality. Then the linear operators and equations should be developed nonlinearly. They included nonlinear Klein-Gordon equation and Dirac equations, and corresponding Heisenberg equation. The quantum commutation and anticommutation should be developed. This theory included the renormalization, which is the correction of Feynman rules of curved closed loops. We think the interaction equations are nonlinear. Many theories, models and phenomena are all nonlinear, for instance, soliton, nonabelian gauge field, the bag model and so on. The superluminal entangled state, which relates the nonlocal quantum teleportation and nonlinearity, is a new fifth interaction. Moreover, the nonlinear effects exist possibly for some interactions, for single particle, for high energy, and for small space-time, etc. The tests of this nonlinear theory are possible [80].

When wave is developed to nonlinear wave, a mathematical result must be that the linear superposition principle does not hold, which should include quantum theory [81-83,19,87]. If the wave-particle duality is breaking [19], many present theories and corresponding conclusions will be corrected and developed. Further, as Bell pointed out, we have an apparent incompatibility, at the deepest level, between the two fundamental pillars of contemporary theory [88]. The breaking symmetry of duality, a contradiction between the uncertainty principle and the constancy of light speed [79], the entangled states [89-98], etc., show that the contemporary theories are complete far from.